\documentclass[prl,twocolumn]{revtex4}

\def\EJ{E_{\rm J}}
\def\IJ{I_{\rm J}}

\begin{document}

\title{Tunneling Spectroscopy of Two-level Systems Inside Josephson Junction}
\author{I. Martin$^1$, L. Bulaevskii$^1$, and A. Shnirman$^2$}
\affiliation{$^1$ Los Alamos National Laboratory, Los Alamos, NM 87545, USA \\
$^2$ Institut f\"{u}r Theoretische Festk\"{o}rperphysik
Universit\"{a}t Karlsruhe, D-76128 Karlsruhe, Germany}

\date{\today}

\begin{abstract}
We consider a two-level (TL) system with energy level separation
$\hbar\Omega_0$ inside a Josephson junction. The junction is
shunted by a resistor $R$ and is current $I$ (or voltage $V=RI$)
biased. If the TL system modulates the Josephson energy and/or is
optically active, it is Rabi driven by the Josephson oscillations
in the running phase regime near the resonance
$2eV=\hbar\Omega_0$. The Rabi oscillations, in turn, translate
into oscillations of current and voltage which can be detected in
noise measurements. This effect provides an option to fully
characterize the TL systems and to find the TL's contribution to
the decoherence when the junction is used as a qubit.
\end{abstract}


\maketitle

Josephson junctions are promising candidates for qubits in quantum computing.
However, to ensure long decoherence time, the phase difference across the
junction should be coupled very weakly with other low-energy degrees of
freedom. Meanwhile, even small-area superconductor-insulator-superconductor
(S-I-S) Josephson junctions, as solid-state mesoscopic systems, have many
low-energy degrees of freedom inside amorphous insulating layer. Generally,
those are phonons (i.e., variables with oscillator-type equidistant energy
spectrum) and the two-level (TL) systems (i.e., systems with non-equidistant
energy spectrum so that only excitations from the ground state to the lowest
excited state are involved at a given frequency). Previously it was shown that
optical phonons inside intrinsic Josephson junctions in cuprate layered superconductors
cause anomalies in the DC current-voltage characteristics. Namely, peaks in the
tunneling current at the voltages corresponding to the phonon frequencies,
$V=\hbar \omega_{ph}/2e$, were observed and mechanisms of their coupling with
the junction phase difference were identified, see
Refs.~\onlinecite{Sch,helm1,helm2,maks}.

While low-energy acoustic phonons are coupled weakly with the
phase difference due to small statistical weight and negligible
optical activity, the TL systems may be much more
effective~\cite{Simmonds_PRL04}.  A TL system may be, for example,
an ion having two possible positions inside potential wells with
tunneling between them. These degrees of freedom interact with the
junction phase difference if they modulate the Josephson energy
and/or, if they are optically active. In the latter case they are
coupled to the phase difference, $\phi(t)$ (we measure the phase
difference in units of magnetic flux), via the electric field
inside the junction, ${\cal E}=\dot{\phi}/d$, where $d$ is the
effective thickness of the junction.

Here we propose a method to characterize the TL systems inside
Josephson junctions. Similar to phonons, the TL systems can cause
anomalies in the DC I-V characteristics. But what is more, a TL
system can precess at its Rabi frequency $\Omega_{\rm R}$ when the
Josephson oscillation frequency matches the level splitting,
$\hbar\omega_{\rm J} \equiv 2eV=\hbar\Omega_0$. This is similar to
well known behavior of atom in the presence of a resonant
electromagnetic wave, see Ref.~\onlinecite{Walls_Milburn}.
However, unlike the optical systems, where the Rabi oscillation
are observed as sidebands ($\Omega_0\pm \Omega_{\rm R}$) in the
emission, in our system, the Josephson current  oscillates at the
Rabi frequency itself.

The transport and noise properties of the current and voltage biased Josephson
junctions have been a subject of extensive studies. The techniques range from
the semiclassical Langevin equation~\cite{Koch_VanHarlingen_Clarke_PRL} which
was generalized using the quantum effective action~\cite{ESA84}, to the
infinite series summation~\cite{Grabert_EPL98,Ingold_PRL99}. For $V \gg I_c R$
the Coulomb Blockade regime is realized, where the Cooper pairs tunnel
incoherently and give rise to the shot noise. We will consider this regime for a 
point-like Josephson junction (its size much smaller than Josephson length). 
We will also assume that $2eV \gg k_{\rm B}T$ and that $\omega_{\rm J}$ is higher than
all the frequencies at which we want to measure the voltage noise.

First we consider a junction with the Josephson energy modified by the
interaction with TL system, described by the pseudospin-1/2 operator ${\bf S}$
\begin{equation}
H_J=-E_J(1+{\bf j}\cdot{\bf S})\cos{2\pi\phi}/{\Phi_0},
\label{imp_phase_interaction}
\end{equation}
where $E_J = \Phi_0I_c/2\pi$, $I_c$ is the Josephson critical current without
TL system, and $\Phi_0 = h/(2e)$ is the flux quantum.
Coupling constants ${\bf j}=(j_x,j_y,j_z)$ characterize 
the modulation of the Josephson critical 
current by the TL system.
We assume a small shunt resistance, $R\ll R_N$, where $R_N$ is the junction's
normal state resistance. This ensures that voltage $V$ on the junction can be
small compared with the superconducting gap $\Delta$ in the running phase
regime. The Hamiltonian of the voltage biased system
reads~\cite{Ingold_Nazarov,Grabert_EPL98}
\begin{equation}
H= \frac{q^2}{2C} + H_{\rm R}(Vt-\phi) + H_J(\phi) - \hbar\Omega_0 S_z\ ,
\end{equation}
where $H_{\rm R}(\xi)$ is the Hamiltonian of the resistor on
which the phase $\xi$ drops, and $q$ is the charge on the capacitor $C$,
$q$ being conjugate to $\phi$. 
After the transformation $\tilde \phi = \phi - Vt$ and $\tilde H = H - V q$ we obtain
\begin{equation}
\label{eq:Hamiltonian_tilde} \tilde H=  \frac{\tilde
q^2}{2C}+ H_{\rm R}(-\tilde\phi) + H_J(\tilde\phi+Vt) - \hbar\Omega_0 S_z\ ,
\end{equation}
where $\tilde q \equiv q-CV$.  The Josephson current operator is
\begin{equation}
\IJ \equiv I_c (1+{\bf{j}\cdot \bf{S}}) \sin\frac{2\pi(\tilde\phi+Vt)}{\Phi_0}
\ .
\end{equation}

For frequencies such that $\hbar \omega \ll 2eV$ and in the running phase regime, $V \gg I_c R$, 
the $symmetrized$ phase autocorrelator $S_{\phi}(\omega)$ is related to the $symmetrized$
current autocorrelator, $S_{I}(\omega)$, via
\begin{equation}
\label{eq:S_I_to_Sphi} S_{\phi} \approx S_{\phi,0} + \frac{R^2}{\omega^2 Y(\omega)}\,S_{I},
\ \ S_{\phi,0} = \frac{\hbar R}{\omega Y(\omega)}\,
\coth\frac{\hbar\omega}{2k_{\rm B}T}\ ,
\end{equation}
where $Y(\omega) \equiv 1+C^2R^2\omega^2$. In Eq.~(\ref{eq:S_I_to_Sphi}) 
the equilibrium correlator 
$S_{\phi,0}$ is due to the first two terms of 
Hamiltonian (\ref{eq:Hamiltonian_tilde}), while the next ($S_I$) term 
accounts for the Josephson coupling. Eq.~(\ref{eq:S_I_to_Sphi}) can be obtained either
from an exact relation between the phase and current Green's functions or
from the quasiclassical Langevin equation.
In typical experiments, $RC\omega \ll 1$ for all relevant frequencies $\omega$,
and, therefore, $Y(\omega) \approx 1$. We, however, keep the factor $Y(\omega)$
to be able to discuss the regime of a junction shunted by a big  capacitor.

Next, we calculate the correlator of the Josephson current operators. We are
particularly interested in a near-resonance situation,  $\Omega_0 \approx
\omega_{\rm J}$. We will use the full quantum Hamiltonian (\ref{eq:Hamiltonian_tilde}) 
to calculate the spin's contribution to the correlator $S_{I}$.  As we can see from
Eq.~(\ref{eq:Hamiltonian_tilde}), the spin is subject to an ac driving at
frequency $\omega_{\rm J}$, ``broadened" by the fluctuating phase $\tilde
\phi(t)$.  Thus it is convenient to transform to the frame rotating with the
angular velocity $(2\pi/\Phi_0)[V + (d/dt)\tilde\phi]$. Formally this amounts
to performing canonical transformation $H' = U \tilde H U^{-1} +i\hbar \dot
UU^{-1}$, with
\begin{equation}
U=\exp\left[\frac{2\pi i}{\Phi_0}\,(\tilde \phi +
Vt)\,S_z\,\right] \ .
\end{equation}
Without lost of generality we take ${\bf
j}=(j_\perp,0,j_\parallel)$. The result is
\begin{eqnarray}
\label{eq:RF_Hamiltonian} H'= && \frac{\tilde q^2}{2C} +H_{\rm
R}(-\tilde\phi)
- \hbar(\Omega_0 - \omega_{\rm J}) S_z - \frac{2e \tilde q S_z}{C}
\nonumber \\
&&  -\EJ(1+j_\parallel S_z) \cos\frac{2\pi(\tilde\phi+Vt)}{\Phi_0} -
\hbar\Omega_{\rm R} S_x\ ,
\end{eqnarray}
where $\Omega_{\rm R} \equiv j_\perp I_c/(4e)$ is the Rabi frequency of the
spin. The counter-rotating term
($\propto\exp {\,\pm\,4\pi(\tilde\phi+Vt)}/{\Phi_0}$)
can be shown to be not important.  The resonance is reached when $\omega_{\rm
J} = \Omega_0$. Then the spin rotates around the $x$-axis (of the rotating
frame) with the Rabi frequency $\Omega_{\rm R}$, but its dynamics is affected
by the noise due to the charge and phase fluctuations.

The operator of the charge on the capacitor
is transformed in the rotating frame as
\begin{equation}
\label{phys_q}
q'=U\tilde q U^{-1} = \tilde q - 2e S_z \ .
\end{equation}
The Josephson current operator transforms as
\begin{eqnarray}
\label{eq:IJ_rotating} &&\IJ'=U\IJ U^{-1}\nonumber \\
&&= I_c (1+j_\parallel S_z)\sin\frac{2\pi(\tilde\phi+Vt)}{\Phi_0}
-\frac{j_\perp I_c}{2}\, S_y
\ ,
\end{eqnarray}
which shows that dynamics of the spin translates into dynamics of the current.

For $|{\bf j}|\ll 1$, the spin's contribution to the average current is
negligible (unless spin's energy relaxation is much faster than that of the
Cooper pairs). This, however, is not always the case for the current near the Rabi frequency 
as described by the correlator $S_{I}(\omega)
\equiv \langle \IJ(t)\IJ(t')\rangle_{\omega}$ at  $\omega \approx \Omega_{\rm R}$. 
Indeed, as we shall see, the
spin-dependent part of $\IJ$ gives rise to a singular (peaked) contribution to
the correlator $S_I$. We calculate $S_I(\omega)$ in the rotating frame using
Eqs.~(\ref{eq:RF_Hamiltonian}) and (\ref{eq:IJ_rotating}). The smooth part of
$S_I(\omega)$ is dominated by the shot noise of the Cooper pairs which tunnel 
incoherently. The lowest order in $RI_c/[VY(\omega)]$ approximation gives good results
for $S_I^{\rm shot}$ if $VY(\omega) \gg I_c R$, $\omega \ll \omega_{\rm J}$ and
$R \ll R_Q\equiv h/4e^2$. Then $S_I^{\rm shot}(\omega) = (I_c^2/4)[P(\omega +
\omega_{\rm J}) + P(\omega - \omega_{\rm J})]$, where $P(t) = \exp[J(t)] $, and
$J(t) =(2\pi/\Phi_0)^2 \langle (\tilde\phi(t)-\tilde\phi(0))\,
\tilde\phi(0)\rangle$ ~\cite{P(E)_Devoret}. At high voltages $2eV \gg k_{\rm B}
T$ we obtain for symmetrized correlator
\begin{equation}
\label{Eq:IcIc}
S_I^{\rm shot}(\Omega_{\rm R}) \approx S_I^{\rm shot}(0)
\approx \frac{I_c^2}{4} P(\omega_{\rm J})\approx \frac{e R I_c^2}{V
Y(\omega_{\rm J})} \ .
\end{equation}

In addition, there exists a peak-shaped contribution to $S_I(\omega)$ near
$\omega \approx \Omega_{\rm R}$. Using
$\langle\sin{2\pi(\tilde\phi+Vt)}/{\Phi_0}\rangle = R I_c/[2VY(\omega_{\rm
J})]\equiv I_{av}$ we take into account the $j_\parallel$ term by defining 
$S_* \equiv j_\perp S_y + (I_{av}/I_c)j_\parallel S_z$.
Then
\begin{equation}
S_I^{\rm spin}(t,t') = \frac{I_c^2}{8}\langle \{S_*(t) S_*(t')\}_+ \rangle \ .
\end{equation}
To calculate this
correlator we use Hamiltonian (\ref{eq:RF_Hamiltonian}). Exactly at resonance
the effective magnetic field $\hbar \Omega_{\rm R}$ is directed along the $x$
axis.  Assuming the Rabi oscillations are under-damped (to be checked for
self-consistency) we obtain
\begin{equation}
S_I^{\rm spin}=\frac{j_{\rm eff}^2
I_c^2}{16}\left[\frac{\Gamma_2} {(\omega-\Omega_{\rm R})^2 +
\Gamma_2^2}+(\omega \rightarrow -\omega) \right] \ ,
\end{equation}
where $j_{\rm eff}^2\equiv j_\perp^2 + (I_{av}/I_c)^2 j_\parallel^2$.

To calculate the dephasing rate $\Gamma_2$ we note that in (\ref{eq:RF_Hamiltonian}) both
the voltage noise $\delta V \equiv \tilde q/C$ and the shot noise (the
$j_\parallel$ term) are coupled to $S_z$, i.e., transversely to the Rabi field.
Thus they contribute to $\Gamma_2$ through the longitudinal relaxation rate $\Gamma_1$, with
$\Gamma_2=(1/2)\Gamma_1$.

The voltage noise consists of two (uncorrelated to the lowest order)
contributions: the equilibrium Johnson-Nyquist one and the one due to the shot
noise of the Cooper pairs. Using $\delta V = (d/dt) \tilde \phi$ (when no spin
is present) and the expression for $S_{\phi,0}$, we obtain for the
Johnson-Nyquist noise
\begin{equation}
S_V^{\rm JN}(\Omega_{\rm
R})=\frac{\hbar \Omega_{\rm R} R}{Y(\Omega_{\rm R})}
\coth\frac{\hbar\Omega_{\rm R}}{2k_{\rm B}T}\ .
\end{equation}
The corresponding rate is $\Gamma_2^{\rm JN}= (e/\hbar)^2 S_V^{\rm
JN}(\Omega_{\rm R})$.

The shot noise contributes twice: as part of the voltage noise
$\delta V$ and via the $j_\parallel$ term in
(\ref{eq:RF_Hamiltonian}). Using Eq.~(\ref{eq:S_I_to_Sphi}) for
the $\delta V$ part we obtain
\begin{equation}
\label{eq:Sq_shot} S_V^{\rm shot}(\Omega_{\rm R}) =
\frac{R^2}{Y(\Omega_{\rm R})} S_I^{\rm shot}(\Omega_{\rm R}) \ .
\end{equation}
The corresponding rate is $\Gamma_2^{{\rm shot},\delta V}=
(e/\hbar)^2 S_V^{\rm shot}(\Omega_{\rm R})$. Finally the
$j_\parallel$ term contributes the rate
\begin{equation}
\Gamma_2^{\parallel} = (j_\parallel/4e)^2 S_I^{\rm
shot}(\Omega_{\rm R})\ .
\end{equation}
The last equation follows from 
$\langle\langle\sin\varphi(t)
\sin\varphi(t')\rangle\rangle=
\langle\langle\cos\varphi(t)\cos\varphi(t')\rangle\rangle$, where 
$\varphi(t)\equiv
2\pi(\tilde\phi(t)+Vt)/\Phi_0$. One can check that all three
rates are to be added ``incoherently", that is the noise cross-terms vanish. In
particular, the cross-term between the two shot noise contributions vanishes
because $\langle\langle\sin\varphi(t)\cos\varphi(t')\rangle\rangle=0$. Thus,
accounting for intrinsic TL dephasing $\Gamma_0$, 
\begin{equation}
\Gamma_2 = \Gamma_2^{\rm JN} + \Gamma_2^{{\rm shot},\delta V} +
\Gamma_2^{\parallel} + \Gamma_0\ .
\end{equation}

For the ``signal", i.e., the height of the voltage peak we obtain
\begin{equation}
\label{eq:height}
S_V^{\rm peak}(\Omega_{\rm R})=\frac{j_{\rm eff}^2 R^2
I_c^2}{16\Gamma_2 Y(\Omega_{\rm R})} \ , 
\end{equation}
while the ``noise", i.e., the smooth background is given by
\begin{equation}
S_V^{\rm bg}=\frac{R^2}{Y(\Omega_{\rm R})} \left( S_I^{\rm
shot}+\frac{\hbar\Omega_{\rm R}}{R}\,\coth\frac{\hbar\Omega_{\rm
R}}{2k_{\rm B}T} \right)\ .
\end{equation}
Collecting all the terms we obtain the signal-to-noise ratio
\begin{eqnarray}
\label{eq:R}
{\cal R} =
\frac{Y(\Omega_{\rm R})
A_\parallel\,\left[\coth\frac{\hbar\Omega_{\rm R}}{2 k_{\rm
B} T} + \frac{e R^2 I_c^2}{\hbar \Omega_{\rm R} V Y(\omega_{\rm
J})}\right]^{-1}}
{\left[\coth\frac{\hbar\Omega_{\rm
R}}{2 k_{\rm B} T} + \frac{e R^2 I_c^2 B_\parallel}{\hbar \Omega_{\rm R} V
Y(\omega_{\rm
J})}+ \frac{2 R_Q Y(\Omega_{\rm R})\Gamma_0}{\pi R\Omega_{\rm R}}\right]}
\ ,
\end{eqnarray}
where $A_\parallel \equiv 1 + \left(\frac{j_\parallel}{j_\perp}\frac{R I_c}{V
Y(\omega_{\rm J})}\right)^2$, $B_\parallel \equiv 1 +Y(\Omega_{\rm R})
\left(\frac{j_\parallel}{2\pi}\frac{R_Q}{R}\right)^2$, and 
$R_Q\equiv h/(4e^2)$.

For purely transverse coupling, $j_\parallel = 0$, the essential
physics is the following: we illuminate the spin with the
``magnetic'' field $2\hbar \Omega_{\rm R} \cos 2\pi(Vt +
\tilde\phi)/\Phi_0$. This field can be thought of as having a
sharp peak (a line) 
near $\omega = \omega_{\rm J}$. The width of this Josephson line is given by the
total voltage noise at zero frequency, $\hbar\Delta \omega =\pi
S_V(0)/R_{\rm Q}=(\pi/R_{\rm Q})(S_V^{\rm JN}(\omega=0) + R^2
\,S_I^{\rm shot}(\omega=0))$.  This relation between the width of the Josephson line 
and the total voltage noise was obtained in
Refs.~\cite{Likharev_Semenov_JETP72,Levinson_PRB03}. The Rabi
oscillation produced by this ``line'' are, in turn, also broadened
by the same amount (in addition to the intrinsic broadening) 
$\Gamma_2 = \Delta\omega/2 + \Gamma_0 $. Finally, the spin's 
(broadened) Rabi precession leads to broadened oscillations of
the Josephson current and voltage at $\Omega_{\rm R}$ on top of the background 
of the JN and shot noise.

It is also important
to note that the Rabi oscillations of the pseudospin correspond to exactly
one Cooper pair going back and forth across the junction. This can be
seen from Eq.~(\ref{phys_q}), or from the fact that for the Rabi oscillations
to occur exactly one ``Josephson photon'' with the energy $\hbar\omega_{\rm J}$
must be absorbed and reemitted by the spin, i.e., exactly one Cooper
pair must go through the voltage drop $V$.

To get a feeling for the relevant numbers we take the data obtained by
Simmonds {\it et al.} \cite{Simmonds_PRL04}, where the two lowest levels
of a junction (phase qubit) in the superconducting (phase-non-running) regime
were driven resonantly. The level splitting $\omega_{01}(I)$ was varied
by the bias current $I$  in the frequency interval
8.6-9.1 GHz. At some values of $\omega_{01}(I)$ appreciable splittings (avoided
level crossings) were observed. This was suggested to originate from TL systems
with $\Omega_0 \approx \omega_{01}(I)$, and the interaction with the phase
difference of the type (\ref{imp_phase_interaction}). 
The splitting is caused by the $j_\perp$ term, while $j_\parallel$ term 
is inessential as long as $j_\parallel \EJ \ll \hbar\Omega_0$. 
Thus, the strongest impurity had $j_\perp
\approx 6.5\cdot 10^{-5}$, while we have an upper bound for the 
strength of the longitudinal coupling $j_\parallel < 10^{-3}$.
This gives $\Omega_{\rm R} \approx 2\pi \times 200\ 
{\rm MHz}$ (the spilitting of $25$ MHz in \cite{Simmonds_PRL04} is due to the
reduction factors corresponding to the zero-point motion of the phase degree of
freedom in the potential well). In Ref.~\cite{Simmonds_PRL04} 
the critical current is $I_c \approx 10\ \mu$A, the normal resistance of the junction 
is $R_N \sim 30\ \Omega$, while  $C \sim 1$ pF. Using these parameters we 
estimate the signal strength and the signal-to-noise ratio ${\cal R}$.
For the temperature we
assume $T \approx 10$ mK, or $k_{\rm B} T/\hbar = 2\pi\times 200$ MHz. The minimum
voltage is given by $I_c R$.  We assume the shunting
resistance of order $R\sim 0.1\Omega \ll R_N$. Shunts of this magnitude have been used
in ~\cite{Koch_VanHarlingen_Clarke_PRB}. Hence, we have
$\omega_{\rm J} > (2e/\hbar) I_c R \approx 2\pi \times 0.5\  {\rm GHz}$. From
above $\omega_{\rm J}$ is restricted by the gap which gives $\omega_{\rm J} <
(2e/\hbar) I_c R_N \approx 2\pi \times 150$ GHz.
Thus we can take $\omega_{\rm J} \sim 2\pi \times 10$ GHz to be in resonance with
the observed TLS. For $j_\parallel=0$ and assuming $\Gamma_0=0$ we obtain the 
signal-to-noise ratio
${\cal R} \approx 0.25$. For the maximally allowed
$j_\parallel=10^{-3}$ the ratio ${\cal R}$ does not change considerably. 

For the integrated signal (signal amplitude) we obtain $\left[\int_{\Omega_{\rm R}} 
\frac{d\omega}{2\pi} S_V^{\rm peak}\right]^{1/2} \approx \frac{j_{\rm eff} R
I_c}{4 \sqrt{2Y(\Omega_{\rm R})}}\approx 10^{-2}$ nV. The Rabi line width
is dominated by the Johnson-Nyquist noise, $\Gamma_2 \approx 2\pi\times 5$ kHz.
If $\Gamma_0$ exceeds  considerably 
this value the signal-to-noise ratio ${\cal R}$ will be reduced.

We also note that for the above introduced parameters we have $1/(CR) \sim 10^{13}\, s^{-1}$. 
Thus $ Y(\Omega_{\rm R}) \approx Y(\omega_{\rm
J}) \approx 1$. One has to increase $C$ by at least three order of magnitude in
order to start having $Y(\Omega_{\rm R}) > 1$.

For some (extreme) parameters, such that $Y(\Omega_{\rm R}) \gg 1$,
we obtain ${\cal R}\gg 1$ which is in contrast
with the limitation ${\cal R}\leq 4$ found for the measurements of the peak in
the current noise at the frequency $\Omega_0$
\cite{Korotkov_Averin_PRB01,Shnirman_EPL04,Bulaevskii_Ortiz_PRB03} using a
normal-state tunnel junction (broad band amplifier). In that case the voltage
$V\gg\hbar\Omega_0/e$ (broad band) is applied. It {\em incoherently} excites
the TL system but also introduces the relaxation due to dissipation necessary
for measurement procedure.  The relaxation is determined by the noise at
frequency $\Omega_0$ and the signal is measured on the background of the noise
at the same frequency. As a result, ${\cal R}$ is a universal number.  In the
case considered here, spin is excited at (high) frequency $\Omega_0$, but the
signal is observed  at low frequency, due to nonlinearity of coupled spin and
Josephson junction system.  It is worth mentioning, however, that in the regime
with large signal-to-noise ${\cal R}$ the single Cooper pair mainly
charges and discharges the capacitor, barely going through the
resistor; thus, the integated signal in this regime is reduced.

Now let's consider the mechanism where spin couples to the junction via the
electric field. The Hamiltonian is
\begin{equation}
\label{eq:H_q}
H=  \frac{q^2}{2C} + H_{\rm R}(Vt-\phi) - \EJ
\cos\frac{2\pi\phi}{\Phi_0} - \hbar\Omega_0 S_z - \frac{Q_{\rm TL}\, q}{C}\,S_x 
\ ,
\end{equation}
where $Q_{\rm TL}$ is the effective charge of the TL system, given by
$Q_{\rm TL} = d_{\rm TL}/L$, where $d_{\rm TL}$ is the TL's dipole
moment, while $L$ is the junction's width. For simplicity we assumed purely
transverse coupling. Remarkably, the splitting observed in  
Ref.~\cite{Simmonds_PRL04} could also be explained by (\ref{eq:H_q}) 
with $Q_{\rm TL} \sim e$. In this mechanism the spin is
coupled to the variable $q$, which in zeroth order in tunneling ($\EJ$) does
not oscillate and has only the Johnson-Nyquist noise spectrum. At $V>R I_c$
this variable acquires an oscillating part due to the Josephson oscillations of
the current. Thus the Rabi driving becomes possible. The width of the Rabi line
is again determined by the full width of the Josephson line $\Gamma_2^{\rm
JN}+\Gamma_2^{\rm shot,\delta V}$. From the integral (weight) of the Josephson peak in
the $S_q(\omega)$ correlator we obtain the Rabi frequency $\Omega_{\rm R} = RI_c
\,Q_{\rm TL}/(2\hbar\sqrt{Y(\omega_{\rm J}}))$. Then, analysis similar to the
one presented above again gives Eq.~(\ref{eq:R}) for the signal-to-noise ratio
(with $A_\parallel=B_\parallel=1$ as we assumed purely transverse coupling).
For $Q_{\rm TL} \sim e$, we obtain a similar to the 
previous mechanism $\Omega_{\rm R}$ and a similar value of ${\cal R}$.  
For the integrated signal we obtain $\left[\int_{\Omega_{\rm R}} 
\frac{d\omega}{2\pi} S_V^{\rm peak}\right]^{1/2} \sim \frac{Q_{\rm TL} R^2
I_c}{e R_Q} \approx 10^{-2}$ nV. Note that the Rabi frequency $\Omega_{\rm R}$ and
the integrated signal depend differently on $R$ in two coupling mechanisms. 
This may allow to distinguish between the two, while they are 
undistinguishable in measurements of type~\cite{Simmonds_PRL04}.

In this letter we discussed what happens when the Josephson oscillations
are in resonance with one TLS. Let us mention another interesting possibility
to manipulate the system. By changing the applied voltage slowly, one can
create the regime of the ``adiabatic passage'' when $\omega_{\rm J}(t)$
passes slowly via $\Omega_0$ and exactly one additional Cooper pair
is transferred through the junction. Varying $\omega_{\rm J}(t)$ in a wide
enough interval one can ``touch'' many TL systems and create a measureable
additional current. 

In conclusion, we propose that the measurements of the low frequency voltage
noise in a Josephson junction in the dissipative (running phase) regime may be
used to characterize the TL systems inside the junctions, i.e., 
energy splitting $\Omega_0$, coupling strength $j_\perp$ from 
the Rabi frequency, and intrinsic dephasing rate $\Gamma_0$ from the 
height of the voltage peak, Eq.~(\ref{eq:height}).  
We predict a peak at
the Rabi frequency when a TL system is resonantly driven by the Josephson
oscillations, $\omega_{\rm J}=\Omega_0$, with the Rabi frequency proportional
to the interaction strength between the TL system and the Josephson phase.  The
peak intensity (signal-to-noise ratio) can be controlled by the shunt resistor
and capacitor.

This work was supported by the
US DOE and by the ESF ``Vortex'' Program. AS acknowledges support
of CFN (DFG) and EU IST Project SQUBIT.

\bibliographystyle{apsrev}
\bibliography{ref}

\begin{thebibliography}{18}
\expandafter\ifx\csname natexlab\endcsname\relax\def\natexlab#1{#1}\fi
\expandafter\ifx\csname bibnamefont\endcsname\relax
  \def\bibnamefont#1{#1}\fi
\expandafter\ifx\csname bibfnamefont\endcsname\relax
  \def\bibfnamefont#1{#1}\fi
\expandafter\ifx\csname citenamefont\endcsname\relax
  \def\citenamefont#1{#1}\fi
\expandafter\ifx\csname url\endcsname\relax
  \def\url#1{\texttt{#1}}\fi
\expandafter\ifx\csname urlprefix\endcsname\relax\def\urlprefix{URL }\fi
\providecommand{\bibinfo}[2]{#2}
\providecommand{\eprint}[2][]{\url{#2}}

\bibitem[{\citenamefont{Schlenga et~al.}(1998)\citenamefont{Schlenga, Kleiner,
  M{\"{o}}sle, Scmitt, and M{\"{u}}ller}}]{Sch}
\bibinfo{author}{\bibfnamefont{K.}~\bibnamefont{Schlenga}},
  \bibinfo{author}{\bibfnamefont{R.}~\bibnamefont{Kleiner}},
  \bibinfo{author}{\bibfnamefont{M.}~\bibnamefont{M{\"{o}}sle}},
  \bibinfo{author}{\bibfnamefont{S.}~\bibnamefont{Scmitt}}, \bibnamefont{and}
  \bibinfo{author}{\bibfnamefont{P.}~\bibnamefont{M{\"{u}}ller}},
  \bibinfo{journal}{Phys. Rev. B} \textbf{\bibinfo{volume}{57}},
  \bibinfo{pages}{14518} (\bibinfo{year}{1998}).

\bibitem[{\citenamefont{Helm et~al.}(1997)\citenamefont{Helm, Preis,
  Forsthofer, Keller, Schlenga, Kleiner, and M{\"{u}}ller}}]{helm1}
\bibinfo{author}{\bibfnamefont{C.}~\bibnamefont{Helm}},
  \bibinfo{author}{\bibfnamefont{C.}~\bibnamefont{Preis}},
  \bibinfo{author}{\bibfnamefont{F.}~\bibnamefont{Forsthofer}},
  \bibinfo{author}{\bibfnamefont{J.}~\bibnamefont{Keller}},
  \bibinfo{author}{\bibfnamefont{K.}~\bibnamefont{Schlenga}},
  \bibinfo{author}{\bibfnamefont{R.}~\bibnamefont{Kleiner}}, \bibnamefont{and}
  \bibinfo{author}{\bibfnamefont{P.}~\bibnamefont{M{\"{u}}ller}},
  \bibinfo{journal}{Physica C} \textbf{\bibinfo{volume}{293}},
  \bibinfo{pages}{60} (\bibinfo{year}{1997}).

\bibitem[{\citenamefont{Helm et~al.}(2000)\citenamefont{Helm, Preis, Walter,
  and Keller}}]{helm2}
\bibinfo{author}{\bibfnamefont{C.}~\bibnamefont{Helm}},
  \bibinfo{author}{\bibfnamefont{C.}~\bibnamefont{Preis}},
  \bibinfo{author}{\bibfnamefont{C.}~\bibnamefont{Walter}}, \bibnamefont{and}
  \bibinfo{author}{\bibfnamefont{J.}~\bibnamefont{Keller}},
  \bibinfo{journal}{Phys. Rev. B} \textbf{\bibinfo{volume}{62}},
  \bibinfo{pages}{6002} (\bibinfo{year}{2000}).

\bibitem[{\citenamefont{Maksimov et~al.}(1999)\citenamefont{Maksimov, Arseyev,
  and Maslova}}]{maks}
\bibinfo{author}{\bibfnamefont{E.}~\bibnamefont{Maksimov}},
  \bibinfo{author}{\bibfnamefont{P.}~\bibnamefont{Arseyev}}, \bibnamefont{and}
  \bibinfo{author}{\bibfnamefont{N.}~\bibnamefont{Maslova}},
  \bibinfo{journal}{Sol. St. Com.} \textbf{\bibinfo{volume}{111}},
  \bibinfo{pages}{391} (\bibinfo{year}{1999}).

\bibitem[{\citenamefont{Simmonds et~al.}(2004)\citenamefont{Simmonds, Lang,
  Hite, Nam, Pappas, and Martinis}}]{Simmonds_PRL04}
\bibinfo{author}{\bibfnamefont{R.~W.} \bibnamefont{Simmonds}},
  \bibinfo{author}{\bibfnamefont{K.~M.} \bibnamefont{Lang}},
  \bibinfo{author}{\bibfnamefont{D.~A.} \bibnamefont{Hite}},
  \bibinfo{author}{\bibfnamefont{S.}~\bibnamefont{Nam}},
  \bibinfo{author}{\bibfnamefont{D.~P.} \bibnamefont{Pappas}},
  \bibnamefont{and} \bibinfo{author}{\bibfnamefont{J.~M.}
  \bibnamefont{Martinis}}, \bibinfo{journal}{Phys. Rev. Lett.}
  \textbf{\bibinfo{volume}{93}}, \bibinfo{pages}{035301}
  (\bibinfo{year}{2004}).

\bibitem[{\citenamefont{Walls and Milburn}(1994)}]{Walls_Milburn}
\bibinfo{author}{\bibfnamefont{D.}~\bibnamefont{Walls}} \bibnamefont{and}
  \bibinfo{author}{\bibfnamefont{G.}~\bibnamefont{Milburn}},
  \emph{\bibinfo{title}{Quantum Optics}} (\bibinfo{publisher}{Springer},
  \bibinfo{year}{1994}).

\bibitem[{\citenamefont{Koch et~al.}(1980)\citenamefont{Koch, {V}an Harlingen,
  and Clarke}}]{Koch_VanHarlingen_Clarke_PRL}
\bibinfo{author}{\bibfnamefont{R.~H.} \bibnamefont{Koch}},
  \bibinfo{author}{\bibfnamefont{D.~J.} \bibnamefont{{V}an Harlingen}},
  \bibnamefont{and} \bibinfo{author}{\bibfnamefont{J.}~\bibnamefont{Clarke}},
  \bibinfo{journal}{Phys. Rev. Lett.} \textbf{\bibinfo{volume}{45}},
  \bibinfo{pages}{2132} (\bibinfo{year}{1980}).

\bibitem[{\citenamefont{Eckern et~al.}(1984)\citenamefont{Eckern, Sch{\"o}n,
  and Ambegaokar}}]{ESA84}
\bibinfo{author}{\bibfnamefont{U.}~\bibnamefont{Eckern}},
  \bibinfo{author}{\bibfnamefont{G.}~\bibnamefont{Sch{\"o}n}},
  \bibnamefont{and}
  \bibinfo{author}{\bibfnamefont{V.}~\bibnamefont{Ambegaokar}},
  \bibinfo{journal}{Phys. Rev. B} \textbf{\bibinfo{volume}{30}},
  \bibinfo{pages}{6419} (\bibinfo{year}{1984}).

\bibitem[{\citenamefont{Grabert et~al.}(1998)\citenamefont{Grabert, Ingold, and
  Paul}}]{Grabert_EPL98}
\bibinfo{author}{\bibfnamefont{H.}~\bibnamefont{Grabert}},
  \bibinfo{author}{\bibfnamefont{G.}~\bibnamefont{Ingold}}, \bibnamefont{and}
  \bibinfo{author}{\bibfnamefont{B.}~\bibnamefont{Paul}},
  \bibinfo{journal}{Europhys. Lett.} \textbf{\bibinfo{volume}{44}},
  \bibinfo{pages}{360} (\bibinfo{year}{1998}).

\bibitem[{\citenamefont{Ingold and Grabert}(1999)}]{Ingold_PRL99}
\bibinfo{author}{\bibfnamefont{G.}~\bibnamefont{Ingold}} \bibnamefont{and}
  \bibinfo{author}{\bibfnamefont{H.}~\bibnamefont{Grabert}},
  \bibinfo{journal}{Phys. Rev. Lett.} \textbf{\bibinfo{volume}{83}},
  \bibinfo{pages}{3721} (\bibinfo{year}{1999}).

\bibitem[{\citenamefont{Ingold and Nazarov}(1992)}]{Ingold_Nazarov}
\bibinfo{author}{\bibfnamefont{G.-L.} \bibnamefont{Ingold}} \bibnamefont{and}
  \bibinfo{author}{\bibfnamefont{Y.~V.} \bibnamefont{Nazarov}}, in
  \emph{\bibinfo{booktitle}{Single Charge Tunneling}}, edited by
  \bibinfo{editor}{\bibfnamefont{H.}~\bibnamefont{Grabert}} \bibnamefont{and}
  \bibinfo{editor}{\bibfnamefont{M.~H.} \bibnamefont{Devoret}}
  (\bibinfo{publisher}{Plenum Press}, \bibinfo{address}{N.Y.},
  \bibinfo{year}{1992}), NATO-ASI Series B: Physics.

\bibitem[{\citenamefont{Devoret et~al.}(1990)\citenamefont{Devoret, Esteve,
  Grabert, Ingold, and Pothier}}]{P(E)_Devoret}
\bibinfo{author}{\bibfnamefont{M.~H.} \bibnamefont{Devoret}},
  \bibinfo{author}{\bibfnamefont{D.}~\bibnamefont{Esteve}},
  \bibinfo{author}{\bibfnamefont{H.}~\bibnamefont{Grabert}},
  \bibinfo{author}{\bibfnamefont{G.~L.} \bibnamefont{Ingold}},
  \bibnamefont{and} \bibinfo{author}{\bibfnamefont{H.}~\bibnamefont{Pothier}},
  \bibinfo{journal}{Phys. Rev. Lett.} \textbf{\bibinfo{volume}{64}},
  \bibinfo{pages}{1824} (\bibinfo{year}{1990}).

\bibitem[{\citenamefont{Likharev and Semenov}(1972)}]{Likharev_Semenov_JETP72}
\bibinfo{author}{\bibfnamefont{K.~K.} \bibnamefont{Likharev}} \bibnamefont{and}
  \bibinfo{author}{\bibfnamefont{V.~K.} \bibnamefont{Semenov}},
  \bibinfo{journal}{JETP Lett.} \textbf{\bibinfo{volume}{15}},
  \bibinfo{pages}{442} (\bibinfo{year}{1972}).

\bibitem[{\citenamefont{Levinson}(2003)}]{Levinson_PRB03}
\bibinfo{author}{\bibfnamefont{Y.}~\bibnamefont{Levinson}},
  \bibinfo{journal}{Phys. Rev. B} \textbf{\bibinfo{volume}{67}},
  \bibinfo{pages}{184504} (\bibinfo{year}{2003}).

\bibitem[{\citenamefont{Koch et~al.}(1982)\citenamefont{Koch, {V}an Harlingen,
  and Clarke}}]{Koch_VanHarlingen_Clarke_PRB}
\bibinfo{author}{\bibfnamefont{R.~H.} \bibnamefont{Koch}},
  \bibinfo{author}{\bibfnamefont{D.~J.} \bibnamefont{{V}an Harlingen}},
  \bibnamefont{and} \bibinfo{author}{\bibfnamefont{J.}~\bibnamefont{Clarke}},
  \bibinfo{journal}{Phys. Rev. B} \textbf{\bibinfo{volume}{26}},
  \bibinfo{pages}{74} (\bibinfo{year}{1982}).

\bibitem[{\citenamefont{Korotkov and Averin}(2001)}]{Korotkov_Averin_PRB01}
\bibinfo{author}{\bibfnamefont{A.~N.} \bibnamefont{Korotkov}} \bibnamefont{and}
  \bibinfo{author}{\bibfnamefont{D.~V.} \bibnamefont{Averin}},
  \bibinfo{journal}{Phys. Rev. B} \textbf{\bibinfo{volume}{64}},
  \bibinfo{pages}{165310} (\bibinfo{year}{2001}).

\bibitem[{\citenamefont{Shnirman et~al.}(2004)\citenamefont{Shnirman, Mozyrsky,
  and Martin}}]{Shnirman_EPL04}
\bibinfo{author}{\bibfnamefont{A.}~\bibnamefont{Shnirman}},
  \bibinfo{author}{\bibfnamefont{D.}~\bibnamefont{Mozyrsky}}, \bibnamefont{and}
  \bibinfo{author}{\bibfnamefont{I.}~\bibnamefont{Martin}},
  \bibinfo{journal}{Europhys. Lett.} \textbf{\bibinfo{volume}{67}},
  \bibinfo{pages}{840} (\bibinfo{year}{2004}).

\bibitem[{\citenamefont{Bulaevskii et~al.}(2003)\citenamefont{Bulaevskii,
  Hru\^ska, and Ortiz}}]{Bulaevskii_Ortiz_PRB03}
\bibinfo{author}{\bibfnamefont{L.~N.} \bibnamefont{Bulaevskii}},
  \bibinfo{author}{\bibfnamefont{M.}~\bibnamefont{Hru\^ska}}, \bibnamefont{and}
  \bibinfo{author}{\bibfnamefont{G.}~\bibnamefont{Ortiz}},
  \bibinfo{journal}{Phys. Rev. B} \textbf{\bibinfo{volume}{68}},
  \bibinfo{pages}{125415} (\bibinfo{year}{2003}).

\end{thebibliography}

\end{document}